
%
%
\input harvmac
%
%
%
%
\ifx\answ\bigans
\else
\output={
  \almostshipout{\leftline{\vbox{\pagebody\makefootline}}}\advancepageno
}
\fi
%
%
%

%
%

%
%
\def\UCSD#1#2{\noindent#1\hfill #2%
\bigskip\supereject\global\hsize=\hsbody%
\footline={\hss\tenrm\folio\hss}}
%
%
\def\abstract#1{\centerline{\bf Abstract}\nobreak\medskip\nobreak\par #1}
%
%
%
%
\edef\tfontsize{ scaled\magstep3}
 \tfontsize  \tfontsize
 \tfontsize \font\titlei=cmmi10 \tfontsize
\font\titleis=cmmi7 \tfontsize \font\titleiss=cmmi5 \tfontsize
\font\titlesy=cmsy10 \tfontsize \font\titlesys=cmsy7 \tfontsize
\font\titlesyss=cmsy5 \tfontsize  \tfontsize
\skewchar\titlei='177 \skewchar\titleis='177 \skewchar\titleiss='177
\skewchar\titlesy='60 \skewchar\titlesys='60 \skewchar\titlesyss='60
%
%
%
%
%
\def\inv{^{\raise.15ex\hbox{${\scriptscriptstyle -}$}\kern-.05em 1}}
\def\lbar{{\lower.35ex\hbox{$\mathchar'26$}\mkern-10mu\lambda}} 

%
%
%
%
\def\dsl{\,\raise.15ex\hbox{/}\mkern-13.5mu D} 
\def\delsl{\raise.15ex\hbox{/}\kern-.57em\partial}
\def\Ksl{\hbox{/\kern-.6000em\rm K}}
\def\Asl{\hbox{/\kern-.6500em \rm A}}
\def\Dsl{\hbox{/\kern-.6000em\rm D}} 
\def\Qsl{\hbox{/\kern-.6000em\rm Q}}
\def\gradsl{\hbox{/\kern-.6500em$\nabla$}}
%
%
\def\lspace{\ifx\answ\bigans{}\else\qquad\fi}
\def\lbspace{\ifx\answ\bigans{}\else\hskip-.2in\fi} 
%
%
\def\boxeqn#1{\vcenter{\vbox{\hrule\hbox{\vrule\kern3pt\vbox{\kern3pt
        \hbox{${\displaystyle #1}$}\kern3pt}\kern3pt\vrule}\hrule}}}
%
%
\def\mbox#1#2{\vcenter{\hrule \hbox{\vrule height#2in
\kern#1in \vrule} \hrule}}
%
%
%
%

\def\CI{{\cal I}}   
  \def\CO{{\cal O}}

%
%
%
%
%

%

\def\bar#1{\overline{#1}}

\def\bra#1{\left\langle #1\right|}
\def\ket#1{\left| #1\right\rangle}

\def\darr#1{\raise1.5ex\hbox{$\leftrightarrow$}\mkern-16.5mu #1}

%
%
\def\frac#1#2{{\textstyle{#1\over #2}}} 
%
%
%
%

%
%
%
%

%
%
\def\ltap{\ \raise.3ex\hbox{$<$\kern-.75em\lower1ex\hbox{$\sim$}}\ }
\def\gtap{\ \raise.3ex\hbox{$>$\kern-.75em\lower1ex\hbox{$\sim$}}\ }
\def\gl{\ \raise.5ex\hbox{$>$}\kern-.8em\lower.5ex\hbox{$<$}\ }
\def\roughly#1{\raise.3ex\hbox{$#1$\kern-.75em\lower1ex\hbox{$\sim$}}}
%
%

%

%
\def\np#1#2#3{{Nucl. Phys. } B{#1} (#2) #3}
\def\pl#1#2#3{{Phys. Lett. } B{#1} (#2) #3}

\def\physrev#1#2#3{{Phys. Rev. } {#1} (#2) #3}

\relax

\noblackbox

\def\asl{\hbox{/\kern-.6500em A}}

\def\clebsch#1#2#3#4#5#6{\left(\left.
\matrix{#1&#2\cr#4&#5\cr}\right|\matrix{#3\cr#6}\right)}

\vskip 1.in
\centerline{{\titlefont{Baryon Hyperfine Mass Splittings}}}
\medskip
\centerline{{\titlefont{in Large N QCD}}}
\vskip .3in
\centerline{Elizabeth Jenkins}
\vskip .2in
\centerline{\sl Department of Physics}
\centerline{\sl University of California, San Diego}
\centerline{\sl 9500 Gilman Drive}
\centerline{\sl La Jolla, CA 92093}
\vfill
\abstract{The hyperfine mass splittings
of baryons in large $N$ QCD are proved to be proportional to
${\bf J}^2$.  Hyperfine mass splittings are first allowed at
order $1/N$ in the $1/N$ expansion.  }  \vfill
\UCSD{\vbox{\hbox{UCSD/PTH 93-19}
\vskip-0.1truecm \hbox{hep-ph/9307244}}}{July 1993}
\eject

Consistency requirements of the large $N$ limit of QCD have recently
led to a new quantitative understanding of the interactions of baryons
with pions \ref\dm{R. Dashen and A.V. Manohar,
UCSD/PTH 93-16}\ref\ejone{E. Jenkins, UCSD/PTH 93-17}\ref\rdam{R.
Dashen and A.V. Manohar, UCSD/PTH 93-18}.  The starting
point of this work is the realization that many physical quantities
diverge with $N$ for arbitrary baryon-pion couplings.  Correct large
$N$ behavior requires that the leading order in $N$ contributions to
these quantities cancel exactly.  The condition of exact cancellation
implies relations between baryon-pion couplings at leading order in
$N$.  These large $N$ consistency conditions are extremely predictive;
all baryon-pion couplings are determined by a single coupling constant
in large $N$ \dm\ejone.  The couplings obtained in large $N$ QCD are
identical to the couplings of the large $N$ Skyrme and
non-relativistic quark models \ref\csm{A.V. Manohar,
\np {248}{1984}{19}}\ref\gersak{J.-L. Gervais and B. Sakita, \physrev
{D30}{1984}{1795}}.  Furthermore, the large $N$ baryon-pion couplings
respect light quark spin-flavor symmetry relations.  The $1/N$
correction to these symmetry relations vanishes \rdam, so that violation of the
symmetry relations first occurs at order $1/N^2$.    Constraints on
other parameters of the baryon-pion chiral Lagrangian can also be
obtained in large $N$ \ref\djm{R. Dashen, E. Jenkins and A.V. Manohar,
UCSD/PTH 93-20}.  These parameters also satisfy light quark spin-flavor
symmetry relations at leading order in the $1/N$ expansion.  The
emergence of an effective light quark spin-flavor symmetry for baryon
couplings in large $N$ explains the phenomenological success of these
symmetry relations for baryon couplings.

The purpose of this paper is to study the hyperfine mass splittings
of baryons in large $N$ QCD.  In large $N$, the spectrum of baryon
states for $N_f =2$ light flavors consists of a degenerate tower of
isospin and angular momentum multiplets $(I,J)$ with $I=J$.  For $N$
odd, the tower of baryon states is $(1/2, 1/2)$, $(3/2,3/2)$, $(5/2,
5/2)$, ..., $(N/2, N/2)$.  A consistent large $N$ expansion for baryons
requires that the baryon multiplets are degenerate upto mass
splittings of order $1/N$ \dm\rdam.  In this work, it is shown that
hyperfine mass splittings amongst multiplets in a degenerate tower of
baryon states must satisfy large $N$ consistency conditions.  These
consistency conditions determine all the hyperfine mass splittings in
terms of the mass splitting of the two lowest spin states.
This unique solution of the consistency conditions yields hyperfine
mass splittings which are identical to those produced by the operator
${\bf J}^2$.  Thus, one concludes that to first non-vanishing order in
the $1/N$ expansion, the hyperfine mass splittings are generated by
${\bf J}^2$.  Further, such a mass splitting is first allowed at order
$1/N$.  These large $N$ results are reminiscient of the Skyrme model
\ref\adw{G.S. Adkins, C.R. Nappi and E. Witten, \np {228}{1983}{552} }
where hyperfine mass splittings are given by ${\bf J}^2/ 2 \CI$, and
the moment of inertia $\CI$ of the baryon states is $\CO(N)$.

The above results can be generalized to baryons which contain a
single heavy quark.  In the limit $m_Q \rightarrow
\infty$, the low-energy strong interactions of heavy quark baryons
are independent of the heavy quark mass, flavor and spin
\ref\iswis{N. Isgur and M.B. Wise, \pl {232} {1989}{113}, \pl
{237}{1990}{527} }.  Thus,
low-energy pion interactions of heavy quark baryons can be analyzed in
terms of the flavor and angular momentum quantum numbers of the light
degrees of freedom of the baryons in the heavy quark symmetry
limit.  For large $N$, $N$ odd, and for $N_f =2$ light flavors, the
light degrees of freedom of baryons containing a single heavy quark
consists of the degenerate tower of $(I,J)$ states  $(0,0)$, $(1,1)$,
$(2,2)$, ...., $((N -1)/2, (N -1)/2)$, where $J$ is the angular
momentum of the light degrees of freedom.  The hyperfine mass
splittings of these states are generated by the operator ${\bf J}^2$
in large $N$.  The mass splittings are first allowed at order $1/N$ in
the $1/N$ expansion.  In order to obtain the hyperfine mass splittings
of the heavy baryon states, heavy baryon states
must be constructed from the spin of the heavy quark $S_Q =
\frac 1 2$ and the tower of states $(I,J)$ for the light degrees of
freedom.  The $(0,0)$ state in the tower corresponds to the
spin-$\frac 1 2$ $\Lambda_Q$ baryon, with the spin of the $\Lambda_Q$
determined by the spin of the heavy quark.  All other $(I,J)$ states
in the tower correspond to a degenerate doublet of heavy baryon
multiplets with isospin $I$ and total spin equal to $I \pm \frac 1 2$,
since $J=I$ for the given tower of states.  Thus, for example, the
$(1,1)$ state corresponds to the spin-$\frac 1 2$ $\Sigma_Q$ and the
spin-$\frac 3 2$ $\Sigma_Q^*$.  In the $m_Q \rightarrow \infty$ limit,
the hyperfine mass splittings of the heavy quark baryons are the
same as the hyperfine mass splittings of the light degrees of
freedom\footnote\dag{Note that the heavy quark baryon hyperfine
mass splittings are proportional to ${\bf J}^2 = {\bf I}^2$, where
$J=I$ is the angular momentum of the light degrees of freedom, not
the total spin of the heavy quark baryon.}.  At order $1/N$ and order
$1/m_Q$ in the heavy quark mass expansion, the doublet of heavy baryon
multiplets for each $(I,J)$ state in the tower is no longer
degenerate.  For instance, a $(\Sigma_Q^* - \Sigma_Q)$ mass splitting
is generated at order $1/(N\,m_Q)$.  The factor of $1/N$ for this
splitting is required since only mass splittings which are suppressed
by $1/N$ are consistent with the large $N$ limit \dm. Thus, the
hyperfine mass splittings for heavy quark baryons contain
contributions of order $1/N$ which preserve heavy quark spin symmetry
and contributions of order $1/(N\,m_Q)$ which violate heavy quark spin
symmetry.  The purely $1/N$ splittings are related by the large $N$
consistency conditions described in this work.  The $1/(N\,m_Q)$
splittings are related by large $N$ consistency conditions derived in
Ref.~\ref\js{E. Jenkins, UCSD/PTH 93-20}.  Ref.~\js\ proves that
these splittings are proportional to the operator ${\bf J}\cdot {\bf
S_Q}$.  These results agree with recent calculations performed in the
Skyrme model \ref\glm{Z. Guralnik, M. Luke and A.V. Manohar, \np
{390}{1993}{474} }\ref\jmhyper{E. Jenkins and A.V. Manohar, \pl
{294}{1992}{273} }.  Baryons containing a single heavy quark arise in
the Skyrme picture as heavy quark meson-soliton bound states
\ref\jmw{E. Jenkins, A.V. Manohar and M.B. Wise, \np {396}{1993}{27}
}, where the solitons are the ordinary baryons containing no heavy
quark of the Skyrme model.  This work shows that these successes of
the Skyrme model actually are consequences of large $N$ QCD.

The results of this paper are obtained by studying the
renormalization of baryon masses due to pion loop corrections.
The pion loop correction can be calculated in chiral perturbation
theory.  The correction to a baryon hyperfine splitting is of the form
\eqn\corr{ \Delta M \rightarrow \Delta M  + \alpha\ {{m_\pi^2}
\over {16 \pi^2 f_\pi^2}} \ln \left( m_\pi^2/ \mu^2 \right)
+ \beta(\mu),
}
where $m_\pi$ is the pion mass, $f_\pi$ is the pion decay
constant, and $\mu$ is a renormalization group subtraction point.  The
$\mu$-dependence of the chiral logarithmic correction is exactly
compensated for by the $\mu$-dependence of the counterterm
$\beta(\mu)$ so that the right-hand side of Eq.~\corr\ is independent
of $\mu$.  The coefficient $\alpha$, which is proportional to
baryon mass splittings, is calculable in chiral
perturbation theory.  Consistency of the large $N$
expansion requires that this coefficient be the same order or higher
order in the $1/N$ expansion as the leading contribution $\Delta M$.

An explicit formula for the coefficient $\alpha$ can be obtained by
studying the chiral logarithmic correction to an individual
baryon mass $M_i$.  The chiral logarithmic correction to a baryon mass
$M_i$ is equal to all one-loop diagrams with a single mass insertion
$M_j$ (summed over intermediate states $j$) minus wavefunction
renormalization $Z_i$ times the tree-level mass $M_i$, as depicted in
\fig\fone{The chiral logarithmic correction to a baryon mass $M_i$. All
baryon states $j$ which are accessible by single pion exchange occur as
intermediate states in the one-loop mass and wavefunction
renormalization diagrams.  Square vertices denote baryon mass
insertions.}. Because the emission of a pion can only change the
isospin or spin of the initial baryon by one unit, the allowed states
$j$ are the $(i-1)$, $i$, and $(i+1)$ states of the baryon tower.
Thus, the chiral logarithmic correction of a baryon mass $M_i$ depends
on the masses $M_{i-1}$, $M_i$, and $M_{i+1}$.  The diagrams
displayed in \fone\ can be reduced to a simpler set of graphs by
noting that a one-loop diagram with a mass insertion can be rewritten
in terms of the diagram without a mass insertion times the mass of the
intermediate baryon, see \fig\ftwo{The one-loop mass insertion diagram
with external state $i$ and intermediate state $j$ is equal to the
mass of the intermediate baryon $M_j$ times the one-loop diagram with
intermediate state $j$ and no mass insertion.}.  This graphical
identity is easily verified using the Feynman rules for baryon
chiral perturbation theory in which the baryon is treated as a
heavy static fermion \ref\jm{E. Jenkins and A.V. Manohar,
\pl{255}{1991}{558}, \pl{259}{1991}{353}}\ref\jmasses{E. Jenkins,
Nucl. Phys. B368 (1992) 190}\ref\jmhungary{E. Jenkins and A.V. Manohar,
{\sl Baryon Chiral Perturbation Theory}, in Proceedings of the
Workshop on ``Effective Field Theories of the Standard Model,'' ed. U.
Meissner, World Scientific (1992) }.
The chiral logarithmic correction to a baryon
mass difference can be obtained by taking the difference of the
chiral logarithmic corrections to each mass.  It is sufficient to
consider hyperfine mass splittings of neighboring multiplets in the
baryon tower.  The chiral correction to the mass difference $(M_{i+1}
- M_{i})$ is given by the diagrams shown in
\fig\fthree{The chiral logarithmic correction to the baryon mass
difference $(M_{i+1} - M_{i})$.  The baryon states $(i-1)$, $i$,
$(i+1)$, and $(i+2)$ are sequential multiplets in the baryon tower,
with $(i-1)$ being the smallest dimensional spin state of the
sequence.  The chiral correction to $(M_{i+1} - M_{i})$ depends
on the mass differences $(M_i - M_{i-1})$, $(M_{i+1} - M_{i})$, and
$(M_{i+2} - M_{i+1})$.  Consistency of the large $N$ limit requires
that this linear combination of mass differences vanishes at leading
order in $N$.}, where $(i-1)$, $i$, $(i+1)$, and $(i+2)$ are
sequential multiplets in the baryon tower, with $(i-1)$ being the
smallest dimensional spin state in the sequence.  The chiral
correction to $(M_{i+1} - M_{i})$ depends on the mass differences
$(M_i - M_{i-1})$, $(M_{i+1} - M_{i})$, and $(M_{i+2} - M_{i+1})$.
The kinematic factors of the one-loop graphs produce the chiral
logarithm in Eq.~\corr\  and are all identical.  Thus, the coefficient
$\alpha$ is proportional to a sum of mass differences times
Clebsch-Gordan factors.  An explicit expression for this linear
combination will be derived below.  Note that \fthree\ assumes that
the baryon mass difference being renormalized is not the mass
difference of the two lowest spin states, so that the state $(i-1)$
exists.  The chiral logarithmic correction to the first mass splitting
is given by the truncated version of \fthree\ shown in \fig\ffour{The
chiral logarithmic correction to the $(M_2 - M_1)$ mass difference.
The correction involves only two mass differences $(M_2 - M_1)$ and
$(M_3~-~M_2)$ since the first state is the smallest spin state
in the baryon tower.}.  In QCD with $N=3$, where the baryon tower
consists of only two multiplets, the chiral logarithmic correction to
the single baryon hyperfine mass splitting $(M_2 - M_1)$ reduces
to the term proportional to $(M_2 - M_1)$ in \ffour\ since there is no
third multiplet in the tower.  This is the renormalization equation
found for the $\Delta -N$ mass difference in Ref.~\jmasses.

The large $N$ power counting for the one-loop graph given in \fthree\
and \ffour\ is determined by the large $N$ behavior of baryon-pion
couplings.  A general baryon-pion coupling is of the form
\eqn\bpib{
\bar B_2 \,G^{ai} B_1 \ {{\partial^a \pi^i} \over f_\pi}\ ,
}
where $a=1,2,3$ labels the angular momentum channel of the $p$-wave
pion, $i=1,2,3$ labels the isospin of the pion, and $G^{ai}$ is an
operator with unit spin and isospin.    Refs.~\dm\ and \ejone\ prove
that the pion couplings amongst baryon multiplets in the degenerate
tower can be parametrized by a single coupling constant $g$,
\eqn\ncouplings{\eqalign{
\bra{ I_2 I_{2z}, J_2 J_{2z} } &G^{ai}
\ket{ I_1 I_{1z}, J_1 J_{1z} } \cr &= N\, g \,
\sqrt{ { 2J_1 +1 } \over { 2J_2 +1} }
\clebsch {I_1}{1}{I_2}{I_{1z}}{i}{I_{2z}}
\clebsch {J_1}{1}{J_2}{J_{1z}}{a}{J_{2z}} \ , \cr
}}
where $I_1=J_1$ and $I_2=J_2$ for the assumed tower of states and
$g$ is $\CO(1)$.  An explicit factor of $N$ has been factored out
of the matrix elements to keep all $N$ dependence manifest.  Note
that since $f_\pi \sim \sqrt{N}$, each baryon-pion coupling
Eq.~\bpib\ produces a net factor of $\sqrt{N}$.  Thus the chiral
logarithmic correction to a baryon mass difference grows with one
additional power of $N$ than a baryon mass difference.
Consistency of large $N$ power counting therefore requires that the
linear combinations of mass differences given in \fthree\ and
\ffour\ vanish identically at leading order in $N$. These constraints
are enough to determine all the baryon hyperfine mass splittings in
terms of one splitting.

Explicit formulae for the large $N$ consistency conditions can be
obtained by evaluating the Clebsch-Gordan factors of the one-loop
graphs in \fthree\ and \ffour.  The Clebsch-Gordan factor for the
one-loop diagram with initial baryon $(I_1, J_1)$ and intermediate
baryon $(I_2, J_2)$ is given by
\eqn\clbidentity{\eqalign{ \sum_{I_{2z},\, J_{2z},\, i,\, a}
&(-1)^i \, (-1)^a
\clebsch{I_1}{1}{I_2}{I_{1z}}{i}{I_{2z}}
\clebsch{J_1}{1}{J_2}{J_{1z}}{a}{J_{2z}}\cr
&\qquad\qquad
\clebsch{I_2}{1}{I_1}{I_{2z}}{-i}{I_{1z}}
\clebsch{J_2}{1}{J_1}{J_{2z}}{-a}{J_{1z}}
= { {(2J_2 + 1)} \over {(2 J_1 +1)} }\ ,\cr
}}
where the factors of $(-1)^i$ and $(-1)^a$ are required since
$(\pi^+)^* = - \pi^-$ in the Condon-Shortley phase convention.
Thus, the constraint that \fthree\ vanishes at leading order in $N$
is
\eqn\recursioni{\eqalign{
(M_{i} - M_{i-1}) { {(2J_{i-1} + 1)} \over {(2J_{i}+1)} }
&- (M_{i+1} - M_{i}) \left[{ {(2J_{i} + 1)} \over {(2J_{i+1}+1)} }+
{ {(2J_{i+1} + 1)} \over {(2J_{i}+1)} }\right]\cr
&+ (M_{i+2} - M_{i+1}) { {(2J_{i+2} + 1)} \over
{(2J_{i+1}+1)} }=0 \ ,\cr  }}
where
\eqn\ji{
J_{i+1} = J_i + 1
}
for the tower of baryon states.  The consistency condition
for the hyperfine mass splitting of the two
smallest dimensional spin multiplets is given by
\eqn\recurinit{
- (M_2 - M_1) \left[{ {(2J_1 + 1)} \over {(2J_2+1)} }+
{ {(2J_2 + 1)} \over {(2J_1+1)} }\right]
+ (M_3 - M_2) { {(2J_3 + 1)} \over {(2J_2+1)} }=0 \ .
}
There is a unique solution to these recursion relations.  Given a
hyperfine mass splitting $(M_2-M_1)$, Eq.~\recurinit\ determines
the next consecutive mass splitting $(M_3 - M_2)$.
Given $(M_2 - M_1)$ and $(M_3 - M_2)$, Eq.~\recursioni\ then
determines the mass splitting $(M_4 - M_3)$.  All remaining mass
splittings are determined recursively using Eq.~\recursioni.

The unique solution to the recursion equations \recursioni\ and
\recurinit\ produces hyperfine mass splittings with the same ratios
as the operator ${\bf J}^2$.  The proof of this assertion is as
follows.  Consider the initial recursion relation Eq.~\recurinit.
This equation fixes the ratio of the first two sequential
hyperfine mass splittings of the baryon tower,
\eqn\ratioone{
{ {(M_3 - M_2)} \over {(M_2 - M_1)} }
= { {(2J_1 +1)(2J_1 +1) + (2J_2 +1)(2J_2 +1) } \over {(2J_1 +1)
(2J_3 +1)} } \ .
}
This result is to be compared with the ratio produced by a mass
contribution proportional to ${\bf J}^2$,
\eqn\ratioone{
{ {(M_3 - M_2)} \over {(M_2 - M_1)} }
= { {J_3(J_3+1) - J_2(J_2+1)} \over {J_2(J_2+1) - J_1(J_1 + 1)} }
={ { J_3 } \over { J_2 } } \
}
since $J_3 = J_2 +1$ and $J_2 = J_1+1$.
These two expressions are not equivalent for general $J_1$.  However,
for the two special cases of interest, baryon towers starting
with $J_1 = \frac 1 2$ or $J_1 =0$, the two expressions are equal.
For these towers, the initial ratios are
\eqn\halfinit{
(M_3 - M_2) = \frac 5 3 \,(M_2 - M_1)
}
for the $J_1 = \frac 1 2$ tower, and
\eqn\zeroinit{
(M_3 - M_2) = 2 \,(M_2 - M_1)
}
for the $J_1 = 0$ tower.
It remains to be shown that given that the initial ratio equals
$J_3/J_2$, all subsequent ratios satisfy
\eqn\ratioj{
{ {(M_{i+2} - M_{i+1})} \over {(M_{i+1} - M_{i})} }= { J_{i+2} \over
J_{i+1} }  \ .
}
The recursion relation Eq.~\recursioni\ gives
\eqn\ratioi{\eqalign{
&{ {(M_{i+2} - M_{i+1})} \over {(M_{i+1} - M_{i})} }=
{ {(2J_{i+1} +1)} \over {(2J_{i+2} +1)} } \ \times \cr
&\qquad\quad
\left\{
\left[  { {(2J_{i+1} +1)} \over {(2J_{i} +1)} } +
{ {(2J_{i} +1)} \over {(2J_{i+1} +1)} }   \right]
-  { {(M_{i} - M_{i-1})} \over {(M_{i+1} - M_{i})} }
{ {(2J_{i-1} +1)} \over {(2J_{i} +1)} }
\right\} \ . \cr
}}
Assuming that the ratio
\eqn\assum{
{ {(M_{i+1} - M_{i})} \over {(M_{i} - M_{i-1})} }
= { J_{i+1} \over J_{i} } \ ,
}
Eq.~\ratioi\ then implies that Eq.~\ratioj\ is satisfied.  This
result is most easily seen by making the substitution $J_i = \frac n
2$.  Then the sought result is equivalent to the identity
\eqn\niden{
{ {(n+3)} \over {(n+5)} }
\left\{
\left[
{ {(n+3)} \over {(n+1)} } + { {(n+1)} \over {(n+3)} }
\right]
 - { {n} \over {(n+2)} } { {(n-1)} \over {(n+1)} }
\right\}
= { {(n+4)} \over {(n+2)} } \ ,
}
which is true for arbitrary $n$.

In summary, this work shows that there is a unique solution of
large $N$ consistency conditions following from pion loop
renormalization of baryon hyperfine mass splittings.  This solution
relates baryon hyperfine mass splittings at leading order in a
$1/N$ expansion.  The solution produces baryon hyperfine mass
splittings proportional to ${\bf J}^2$.
Previous work \dm\rdam\ has proven that mass splittings
are first allowed at order $1/N$.

\vfill\break\eject

\centerline{\bf Acknowledgements}
I thank R.~Dashen and A.V.~Manohar
for discussions, and I thank the Aspen Center for Physics
for hospitality while this work was being completed.
This work was
supported in part by the Department of Energy
under grant number DOE-FG03-90ER40546.

\listrefs
\listfigs
\vfill
\eject

\bye